\documentclass[twocolumn]{aastex631}
\usepackage{bm}

\newcommand{\mytilde}{\raise.19ex\hbox{$\scriptstyle\sim$}}

\usepackage[utf8]{inputenc}
\usepackage{amsmath}
\usepackage{tabularx}
\usepackage{hyperref}
\usepackage{afterpage}

\received{}
\revised{}
\accepted{}

\shorttitle{}

\begin{document}

\title{A High-Caliber View of the Bullet Cluster Through JWST Strong and Weak Lensing Analyses}
\correspondingauthor{M. James Jee}
\email{sang6199@yonsei.ac.kr, mkjee@yonsei.ac.kr}

\author[0000-0001-7148-6915]{Sangjun Cha}
\affiliation{Department of Astronomy, Yonsei University, 50 Yonsei-ro, Seoul 03722, Korea}
\author[0009-0009-4273-6132]{Boseong Young Cho}
\affiliation{Department of Astronomy, Yonsei University, 50 Yonsei-ro, Seoul 03722, Korea}
\author[[0000-0001-9139-5455]{Hyungjin Joo}
\affiliation{Department of Astronomy, Yonsei University, 50 Yonsei-ro, Seoul 03722, Korea}
\author[0000-0002-1566-5094]{Wonki Lee}
\affiliation{Department of Astronomy, Yonsei University, 50 Yonsei-ro, Seoul 03722, Korea}
\author[0000-0002-2550-5545]{Kim HyeongHan}
\affiliation{Department of Astronomy, Yonsei University, 50 Yonsei-ro, Seoul 03722, Korea}
\author[0009-0009-4086-7665]{Zachary P. Scofield}
\affiliation{Department of Astronomy, Yonsei University, 50 Yonsei-ro, Seoul 03722, Korea}
\author[0000-0002-4462-0709]{Kyle Finner}
\affiliation{IPAC, California Institute of Technology, 1200 E California Blvd., Pasadena, CA 91125, USA}
\author[0000-0002-5751-3697]{M. James Jee}
\affiliation{Department of Astronomy, Yonsei University, 50 Yonsei-ro, Seoul 03722, Korea}
\affiliation{Department of Physics and Astronomy, University of California, Davis, One Shields Avenue, Davis, CA 95616, USA}

\begin{abstract}
The Bullet Cluster (1E 0657-56) is a key astrophysical laboratory for studying dark matter, galaxy cluster mergers, and shock propagation in extreme environments. Using new JWST imaging, we present the highest-resolution mass reconstruction to date, combining 146 strong lensing constraints from 37 systems with high-density (398 sources arcmin$^{-2}$) weak lensing data, without assuming that light traces mass. The main cluster's mass distribution is highly elongated (NW-SE) and consists of at least three subclumps aligned with the brightest cluster galaxies. The subcluster is more compact but elongated along the E-W direction, with a single dominant peak. We also detect a possible mass and ICL trail extending from the subcluster's eastern side toward the main cluster. Notably, these detailed features are closely traced by the intracluster light, with a modified Hausdorff distance of $19.80 \pm 12.46$ kpc.
Together with multi-wavelength data, the complex mass distribution suggests that the merger history of the Bullet Cluster may be more complex than previous binary cluster merger scenarios.
\end{abstract}

\section{Introduction} \label{sec:intro}

\begin{figure*}
\centering
\includegraphics[width=0.95\textwidth]{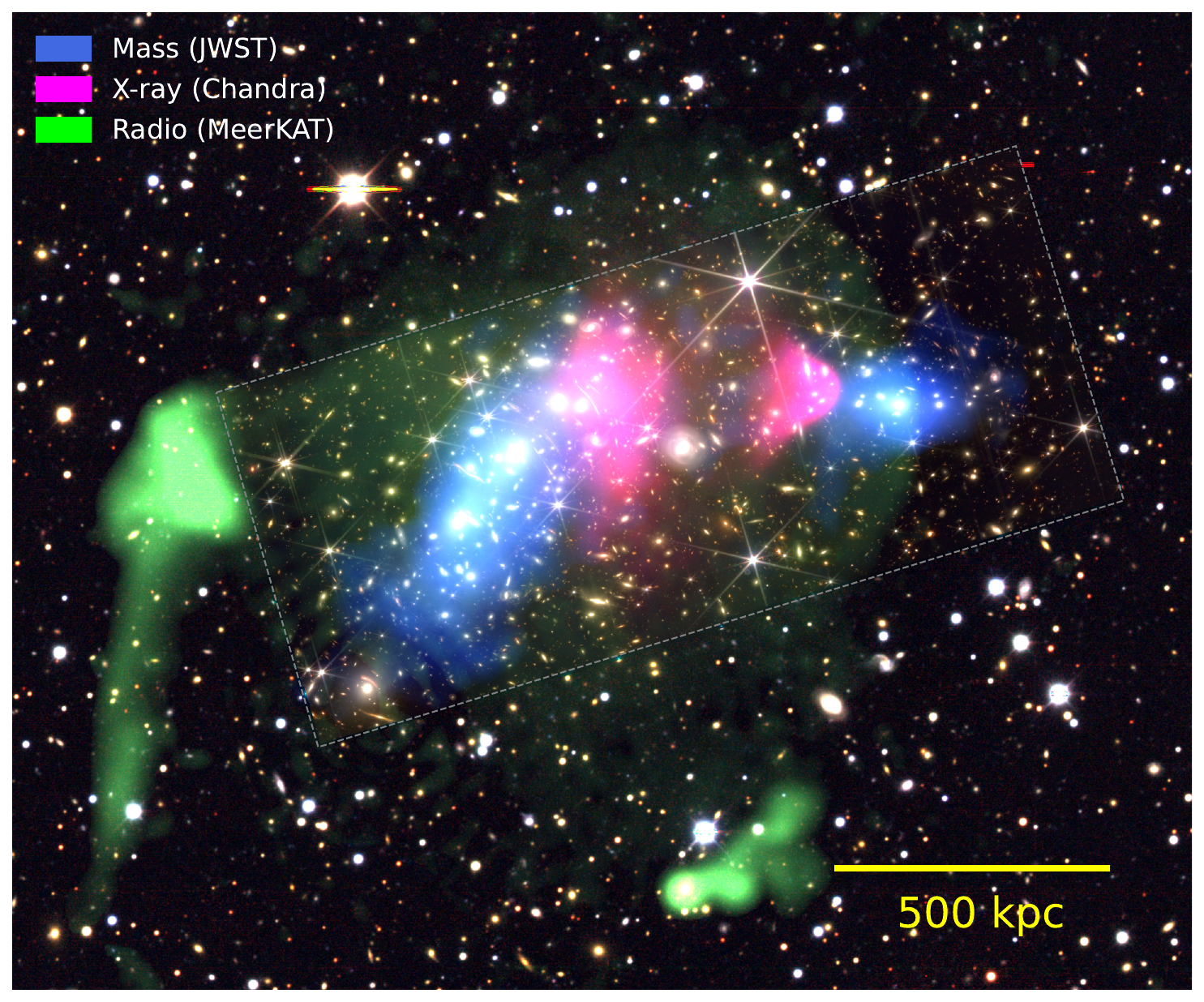} 
\caption{Multi-wavelength view of the Bullet Cluster. North is up and east is left. The background color-composite image, covering the  $2.2~\mbox{Mpc}\times1.8~\mbox{Mpc}$ region, was created using DECam data, with the 
$g$, $r$, and $i$ bands mapped to blue, green, and red intensities, respectively.
The rectangular box outlined with a gray dashed line marks the JWST color-composite image, constructed using F090W+F115W+F150W, F200W+F277W, and F356W+F410M+F444W as the blue, green, and red channels, respectively.
The mass distribution of the main cluster reconstructed with the JWST data is highly elongated in the northwest-southeast direction and consists of at least three subclumps, each coinciding with the
brightest cluster galaxies. In contrast, the mass distribution of the subcluster is relatively compact but elongated in the east-west direction, with a single dominant mass peak centered at the brightest galaxy. The western edge of the MeerKAT emission traces the ``Bullet"-side shock front, which is also detected as a surface brightness jump in the X-ray emission. The shock front on the main cluster side is believed to coincide with the giant ($\mytilde1$~Mpc) radio relic extending in the north-south direction.
} 
\label{multi-wavelength}
\end{figure*}

Merging galaxy clusters serve as a cosmic laboratory, offering unique opportunities to study energy scales that are impossible to reproduce on Earth \citep{1999ApJ...521..526M, 2001ApJ...561..621R, 2012ARA&A..50..353K}. The Bullet Cluster (1E 0657-56) is a remarkable astrophysical system, presenting a rare, high-velocity, early-phase merger characterized by a sharp cold front, nicknamed the ``bullet'', a well-defined shock front, and large offsets between the dark matter (DM) and the intracluster medium (ICM) \citep[e.g.,][]{2002ApJ...567L..27M, 2004ApJ...604..596C, 2006ApJ...648L.109C, 2006ESASP.604..723M, 2008MNRAS.384..343N}.

Interestingly, despite its undeniable importance, the Bullet Cluster still remains one of the most challenging systems to reproduce in numerical simulations. For example, while many simulations successfully replicate the observed offset ($\mytilde0.2$ Mpc) between the X-ray emission and DM halo in the subcluster,  none have reproduced the larger offset ($\mytilde0.3$ Mpc) in the main cluster  \citep[e.g.,][]{2007MNRAS.380..911S, 2008MNRAS.389..967M, 2014ApJ...787..144L}. Other challenges include the misalignment between the merger axes inferred from the DM halos and shock features, the extreme collision velocities \citep[e.g.,][]{2007ApJ...661L.131M}, and the survival of the Bullet-like cold front. These challenges have motivated the exploration of the Bullet Cluster as a testbed for the $\Lambda$CDM cosmology and alternative DM theories \citep[e.g.,][]{2008ApJ...679.1173R, 2010ApJ...718...60L, 2017MNRAS.465..569R}.

One of the most significant obstacles in numerically reproducing the Bullet Cluster is the challenge of establishing the correct initial conditions prior to the merger. Most existing simulations have assumed binary mergers between virialized halos with initial pre-merger masses inferred from the current (post-merger) data. Cluster mergers involve complex physical processes that significantly modify the structures of both the ICM and DM halos. As a consequence, inferring pre-merger properties from post-merger data is inherently limited. While the effect is more pronounced in the ICM, dark matter halos, the dominant contributor of gravity, also undergo substantial changes due to gravitational interactions during the merger \citep[e.g.,][]{2023ApJ...945...71L}. 

Undoubtedly, the first step toward a robust inference of pre-merger conditions is the precise characterization of the merging system's properties in the current (observed) epoch. In this regard, gaining a deeper understanding of the current dark matter distribution in the Bullet Cluster should be a top priority.

In this study, we present an improved mass model of the Bullet Cluster using new JWST observations, combining both strong lensing (SL) and weak lensing (WL) data. Thanks to the unprecedented depth and resolution, the current analysis benefits from the large number (146) of multiple image positions and the high-density ($\mytilde400$ sources per sq. arcmin) WL data, increasing the number of constraints by several factors compared to previous studies in both SL and WL \citep[e.g.,][]{2006ApJ...648L.109C, 2006ApJ...652..937B, 2009ApJ...706.1201B, 2016A&A...594A.121P}. 
In addition, we present the first JWST study of intracluster light (ICL) in the Bullet Cluster, with an emphasis on its comparison to the DM distribution. As intracluster stars are bound to the cluster's gravitational potential not to individual galaxies, they have been considered visual tracers of the DM distribution \citep[e.g.,][]{2010ApJ...717..420J, 2019MNRAS.482.2838M, 2022ApJS..261...28Y}. The superb imaging quality of the JWST data provides a critical testbed for this hypothesis in this dynamically active environment.

We utilize the free-form method {\tt MARS} \citep{2022ApJ...931..127C, 2023ApJ...951..140C, 2024ApJ...961..186C} to reconstruct the mass distribution. {\tt MARS} provides a lensing data-driven mass map without assuming that light traces mass. This ability is particularly valuable for identifying substructures without relying on the cluster galaxy distribution as an informed prior. Furthermore, it allows us to use the Bullet Cluster to test the self-interacting dark matter (SIDM) hypothesis, as the mass reconstruction algorithm is entirely blind to the baryonic distribution. Together with the ICM and shock tracers from X-ray and radio observations, 
our JWST high-resolution mass and ICL distributions are expected to provide critical insights into the formation of the Bullet Cluster. 

Due to the limited field of view of JWST, the current analysis does not constrain the total mass of the Bullet Cluster without substantial extrapolation beyond the observed region. As such, this paper focuses on the high-resolution mass distribution within the JWST coverage. A comprehensive measurement of the total cluster mass, incorporating wide-field imaging data, will be presented in a forthcoming study (B. Cho et al., in prep).

This paper is outlined as follows. Data and our analysis methods are described in \textsection\ref{sec:data_method}. In \textsection\ref{sec:result}, we present our results and discuss them in \textsection\ref{sec:discussion}. We summarize our conclusion in \textsection\ref{sec:conclusion}.
We assume a flat $\Lambda$CDM cosmology with the dimensionless Hubble constant parameter $h=0.7$ and the matter density $\Omega_{M}=1-\Omega_{\Lambda}=0.3$ unless stated otherwise. The plate scale at the cluster redshift ($z=0.296$) is $4.413 ~\rm kpc ~\rm arcsec^{-1}$.

\section{Data Reduction \& Analysis} \label{sec:data_method}
\subsection{JWST NIRCam Images} \label{optical_data}
We used the publicly available Near Infrared Camera (NIRCam) observations from the JWST program GO-4598 (PI: Maru{\v{s}}a {Brada{\v{c}}}).  
Our data reduction followed the standard JWST pipeline \citep{Bushouse2024}. For calibration, we used the parameter reference file map jwst\_1241.pmap\footnote{\url{https://jwst-crds.stsci.edu/}}. To remove snowball-like artifacts caused by cosmic rays and pink noise from the detector, we adopted the algorithm introduced by \citet{2023ApJ...946L..12B}. We removed the wisp artifacts with the third version of the templates\footnote{\url{https://stsci.app.box.com/s/1bymvf1lkrqbdn9rnkluzqk30e8o2bne}} to minimize the loss of ICL signals. During resampling, we set the output pixel scale to $0\farcs02 ~\rm
 pixel^{-1}$ and selected a square kernel with a {\tt pixfrac} of 0.8.

We identified new SL multiple image candidates through the eight filter-imaging data (F090W, F115W, F150W, F200W, F277W, F356W, F410M, and F444W). For WL shape measurement, the F200W filter is used since it provides the optimal sampling of the point-spread function (PSF) \citep{2023ApJ...953..102F, 2023ApJ...958...33F}. 
We use the F277W image as the primary filter for ICL analysis, as it provides the strongest ICL signal among the long-wavelength channels, which are not affected by chip gaps and are less susceptible to readout noise (\textsection\ref{icl_analysis}).

\subsection{Photometric Redshift Estimation}\label{photoz}
We estimated the photometric redshift using \texttt{EAzY} \citep{2008ApJ...686.1503B} with the \texttt{SFHz\_CORR} templates. In addition to the JWST NIRCam filters, we also obtained photometry from HST/ACS images observed with F435W, F606W, F775W, F814W, and F850LP filters to improve the reliability of the redshift estimation. 

We compared the estimated redshifts with spectroscopic redshifts from previous studies \citep{2017A&A...606A.122F, 2020A&A...634A.137P, 2021A&A...646A..83R} to assess the robustness of our measurement.    
Among the 103 sources matched by coordinates, 86 have photometric redshifts within 10\% of their spectroscopic redshifts. Applying the catastrophic outlier criterion $\left| z_{\text{spec}} - z_{\text{phot}} \right| / (1+z_{\text{spec}}) \geq 0.10$, we identified 17 sources (16.5\%) as outliers.  
The normalized median absolute deviation (NMAD) is $\sigma_{\text{NMAD}} = 0.0012$.

\subsection{SL Data} \label{sl_data}
We compiled SL multiple images from the literature \citep{2016A&A...594A.121P, 2021A&A...646A..83R}  and identified new multiple-image candidates using JWST data. 
The literature-identified images were refined based on improved morphological identification in JWST imaging and updated photometric redshifts.
Using the eight-filter JWST imaging data, we newly identified 52 multiple images from 20 systems (66 positions including internal knots). In addition, we detected new 33 image positions associated with previously known systems. In total, we added 99 new constraints for SL modeling.
The final catalog includes 146 multiple images, where distinct knots within the same source are counted as individual constraints \footnote{The full multiple image catalog can be accessed via \dataset[doi:10.5281/zenodo.15208501]{https://doi.org/10.5281/zenodo.15208501}}.

\subsection{Weak Lensing Analysis} \label{wl_analysis}
We modeled the PSF by applying principal component analysis (PCA) \citep{2007PASP..119.1403J} to the final mosaic image. Since we find that the JWST PSF pattern in the current dataset shows no significant variation across the field, this simplified PSF-modeling approach is justified, as demonstrated in \citet{2023ApJ...953..102F} and \citet{2024ApJ...961..186C}. To test the sensitivity of our results to the PSF model, we produced an alternative PSF model using the \texttt{STPSF} package following the same technique described in \cite{2024ApJ...961..186C}. The resulting shape catalog yielded a SL+WL lens model that was virtually identical to the one obtained with our PCA-based PSF model.

The average residual ellipticity of our PSF model is negligibly small ($8.0\times10^{-6}\pm 2.9\times10^{-3}$ for $e_1$ and $-8.7\times10^{-6}\pm2.3\times10^{-3}$ for $e_2$). In this study, we adopted multiplicative bias factors of 1.11 for $e_1$ and 1.07 for $e_2$, as derived by \citet{2023ApJ...953..102F} through iterative image simulations mimicking the observed field \citep{2013ApJ...765...74J}.

Photometric redshifts ($z_{\rm phot}$) are used to select WL background galaxies. We selected galaxies whose photometric redshift uncertainties are smaller than 25\% and 1$\sigma$ lower bounds are greater than 0.35. This criterion ensures that the line-of-sight velocity difference is six times greater than the velocity dispersion of a typical massive merging cluster \citep{2019ApJ...882...69G, 2025ApJS..277...28F}. Additionally, galaxies brighter than magnitude 22 in the F200W filter were excluded to remove possible contamination from the foreground and cluster members.

For the shape criteria, we discarded the following objects: 1) unstable fitting results\footnote{It is based on the {\tt STATUS} parameter, which is one of the outputs of the {\tt MPFIT} package \citep{2009ASPC..411..251M} for our $\chi^2$ minimization. We considered the fitting to be unstable unless {\tt STATUS} was unity.}, 2) highly elongated shapes (ellipticity greater than 0.9), 3) semi-minor axes smaller than 0.4 pixels, or 4) ellipticity measurement errors greater than 0.3.
Additionally, we discarded objects affected by bright stellar diffraction spikes.
Finally, we manually identified and removed spurious detections, blended sources, over-deblended features, and other imaging artifacts.

\subsection{ICL Analysis} \label{icl_analysis}
For our ICL analysis, we primarily use the F277W filter image for the following reasons. 
First, the long-wavelength channels provide a higher signal-to-noise ratio (SNR) compared to the short-wavelength channels under the same exposure time.
Second, the long wavelength channels are not affected by chip gaps and are less susceptible to readout noise.
Lastly, at the cluster redshift of $z = 0.296$, the F277W filter shows the strongest ICL signal among the long-wavelength filters, as it is sensitive to old stellar populations and benefits from relatively low thermal background noise.

Accurate background estimation is essential for minimizing bias in the measurement of ICL.
The target field in this study contains a large number of bright sources as well as strong ICL emission. 
To measure the background level, we masked all bright sources and analyzed the distribution of the remaining pixel values.
The mode of this distribution was defined as the sky background and subtracted from the data. 
To estimate the surface brightness limit, we calculated the difference between the mode and the 32nd percentile on the negative side of the distribution, as contamination from bright sources may still affect the positive side.

To investigate the spatial distribution of the ICL, we constructed two-dimensional ICL maps for all available JWST filters. 
We obtained a segmentation map using \texttt{SExtractor} \citep{1996A&AS..117..393B} to mask sources, including the brightest cluster galaxy (BCG). 
To further refine the mask, we expanded the mask map proposed by \citet{2023Natur.613...37J} up to the BCG-ICL transition.
The transition radii were determined based on the radial surface brightness profile of each BCG. 
The mask radii for the southern and northern BCGs in the main cluster and the BCG in the subcluster are 56.17, 44.74, and 59.59 kpc, respectively. 
We note that these values are consistent with the range suggested by \citet{2022ApJ...928...99C}.

\begin{figure*}
\centering
\includegraphics[width=0.9\textwidth]{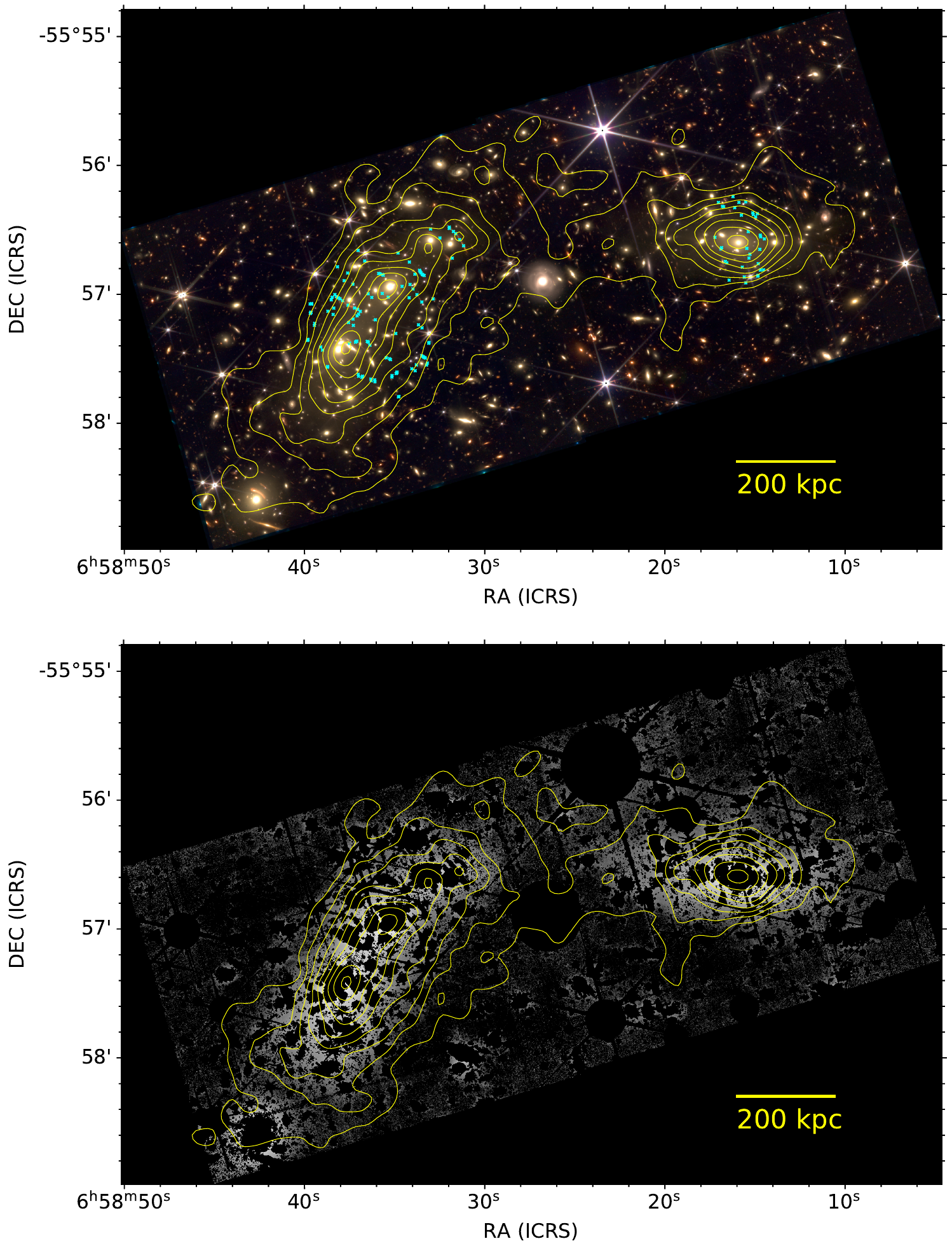} 
\caption{Mass distribution overlaid on the JWST color-composite and ICL map. The contours correspond to $\kappa$ ranging from 0.15 to 1.35 in intervals of 0.15. A Gaussian smoothing with $\sigma = 4\arcsec$ is applied to mitigate pixel-scale artifacts.
(a) Mass contours overlaid on the JWST color image. The cyan crosses mark the positions of all SL multiple images used in this paper. The color-composite image combines filters F444W + F410M + F356W (red channel), F277W + F200W (green channel), and F150W + F115W + F090W (blue channel). 
(b) Mass contours overlaid on the F277W ICL map. All light sources including BCGs are masked out. }
\label{fig:mass_icl_combined}
\end{figure*}

\subsection{Lens Modeling} \label{lens_modeling}
We used the grid-based, free-form lensing reconstruction algorithm {\tt MARS} \citep{2022ApJ...931..127C, 2023ApJ...951..140C} to perform SL+WL analysis. Our minimization target function consists of three terms: two $\chi^2$ terms for SL and WL constraints and one regularization term. The SL term minimizes the scatter of multiple-image positions in the source plane, while the WL term minimizes $\chi^2$ between observed and model-predicted reduced shear. The regularization term maximizes cross-entropy to maintain smoothness and prevent overfitting in mass reconstruction, as demonstrated in previous studies \citep[e.g.,][]{2022ApJ...931..127C, 2023ApJ...951..140C, 2025ApJ...979...13P}. For further details, we refer the reader to \citet{2024ApJ...961..186C}. 

In this study, we employ a uniform $200\times200$ grid covering a $6.4\arcmin\times6.4\arcmin$ (1.7~Mpc$\times$1.7~Mpc) region. For SL sources without spectroscopic redshifts, we treat their redshifts as free parameters, constrained by prior intervals based on the uncertainties from photometric redshift estimates.
For multiple images without any redshift information, and cannot be converged within the prior, we set its prior interval to $z_{model}=[z_{cluster}+0.1, 15]$, as done in \citet{2024ApJ...961..186C}.
Our best-fit mass model predicts the observed multiple-image positions with an RMS scatter of $\Delta_{\rm rms}=0\farcs24$ between the observed and  predicted positions on the image plane. We find that  some multiple-image systems show large discrepancies between the model-predicted and photometric redshifts.
To assess their impact, we repeated our mass reconstruction while fixing the source redshifts at their best-fit photometric redshifts. We verify that the result is highly consistent with the case where the redshifts are allowed to vary.
In our presentation of the convergence $\kappa$, the critical surface mass density  (i.e., the unit of $\kappa$) is $\Sigma_c=1.827\times10^9 ~ M_{\odot} ~ {\rm kpc}^{-2}$, scaled to $D_{\rm ls}/D_{\rm s}=1$.

\section{Results} \label{sec:result}

\subsection{Mass Map}\label{mass_map}

Figures~\ref{multi-wavelength} and~\ref{fig:mass_icl_combined} present the SL+WL mass reconstruction based on the current JWST data. Overall, the mass map reveals two primary mass components of the Bullet Cluster, the main cluster and the subcluster, in agreement with previous studies \citep[e.g.,][]{2006ApJ...648L.109C, 2006ApJ...652..937B, 2009ApJ...706.1201B, 2016A&A...594A.121P}. 
Below, we discuss significant substructures revealed by our free-form mass reconstruction based on the JWST data, which provide several times more lensing constraints than previous analyses. 

In the main cluster, the reconstruction reveals two prominent mass peaks near the two BCGs, separated by $\mytilde$200 kpc.
The mass density between the two BCGs is significantly higher than expected from a simple superposition of two halo profiles, each centered on one of the BCGs. A similar enhancement is also observed in the ICL distribution (\textsection\ref{icl_result}).
The mass distribution in the main cluster extends further to the northwest, closely following the distribution of the brightest cluster galaxies in this region. Overall, the mass distribution in the main cluster is highly elongated along the northwest–southeast direction. In contrast, the subcluster exhibits a more compact mass distribution, though still elongated along the east-west axis, with a single dominant mass peak that aligns well with its BCG. Notably, the mass on the eastern side of the subcluster decreases more gradually than on the western side. Although this trailing feature exhibits low contrast, we find it to be statistically significant and well-correlated with the ICL. A quantitative analysis of this feature is presented in \textsection\ref{icl_result} and \textsection\ref{subcluster_tail}.

The total projected masses within a radius of 250 kpc are estimated to be $M(<250~{\rm kpc}) = 1.59\pm0.14\times10^{14}~M_{\odot}$, $1.67\pm0.14\times10^{14}~M_{\odot}$, and $0.87\pm0.13\times10^{14}~M_{\odot}$ for apertures centered on the northern main cluster BCG, the southern main cluster BCG, and the subcluster BCG, respectively. These values are approximately 50–60\% lower than those reported in previous studies \citep{2006ApJ...652..937B, 2016A&A...594A.121P}. These aperture masses, measured within a relatively small radius, are primarily constrained by our SL data, which offer significantly improved precision. However, because most of our SL constraints rely on photometric redshifts, the exact mass estimates remain subject to revision once spectroscopic redshifts become available.

Despite the superb quality of the JWST lensing data, its limited field of view precludes a robust estimate of the total mass of the Bullet Cluster system without significant extrapolation. This will be addressed in a future study using a joint analysis of JWST and wide-field DECam imaging data (B. Cho et al., in prep).

\subsection{ICL map}\label{icl_result}
\begin{figure}
\centering
\includegraphics[width = \columnwidth]{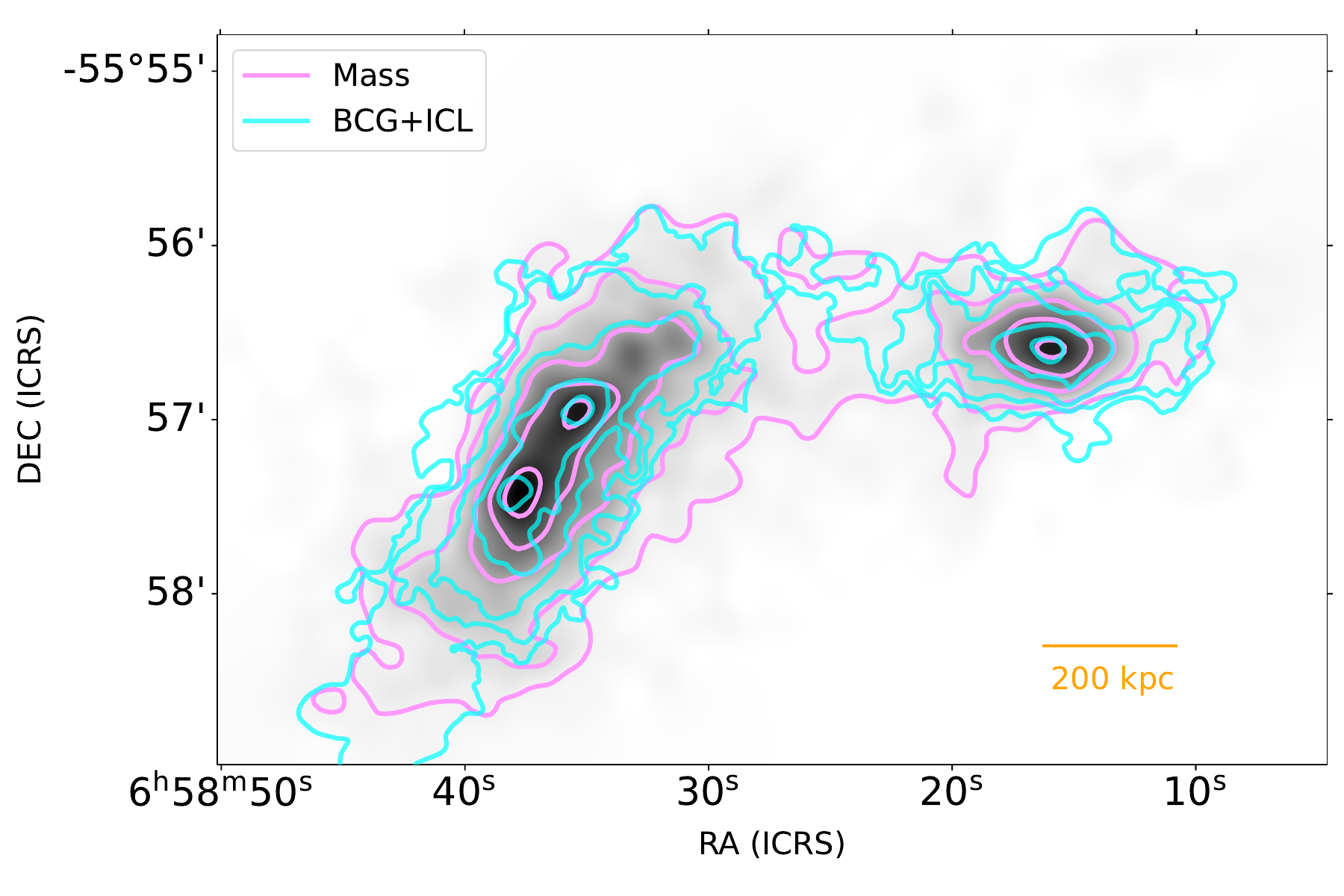}
\caption{Comparison of the SL+WL mass and BCG+ICL surface brightness contours. The pink contours correspond to $\kappa = [1.2, 0.9, 0.6, 0.3, 0.15]$, whereas the skyblue contours correspond to surface brightness levels of $\mu=[2000,500,160,70,10]$ counts per pixel. Using the threshold $r=400~$kpc, the MHD value calculated with these contours is $19.80\pm12.46\ \mathrm{kpc}$, which shows a strong similarity between the two distributions.}
\label{fig:MHD}
\end{figure}

We present a two-dimensional ICL map of the Bullet Cluster in the lower panel of Figure~\ref{fig:mass_icl_combined}. Remarkably, the ICL distribution shows excellent spatial correlation with the mass distribution on both large and small scales. Qualitatively, we note the following:
\begin{itemize} 
\item In the main cluster, the ICL distribution is elongated along the northwest–southeast axis and traces the substructures seen in the mass map.
 \item The ICL map exhibits an overdensity in the region between the two main cluster BCGs, similar to that seen in the mass map (see discussion in \textsection\ref{mass_map}). 
\item In the subcluster, the ICL distribution is elongated along the east–west direction, consistent with features in the mass map. 
\item The ICL on the eastern side of the subcluster decreases more gradually than on the western side, a trend also indicated in the mass reconstruction. We provide a quantitative discussion in \textsection\ref{subcluster_tail}. 
\end{itemize}

For the quantitative comparison between the distributions of DM and BCG+ICL, we calculated the modified Hausdorff distance (MHD). 
The Hausdorff distance is a metric commonly used to quantify the dissimilarity between two data sets, with applications in image analysis, pattern recognition, and computer vision. The MHD is a modified version designed to be more robust to noise and outliers by replacing the maximum operator with an average. Smaller MHD values indicate greater similarity between the two data sets. Readers are referred to \cite{Dubuisson1994} and \cite{Huttenlocher1993} for further details.

Figure~\ref{fig:MHD} shows the reference levels for both the mass and ICL of the Bullet Cluster used in our MHD evaluation. Within a radius of $140$ kpc from each BCG, 
the MHD is estimated to be $15.47\pm11.98\ \mathrm{kpc}$. \cite{2019MNRAS.482.2838M} evaluated MHDs for six Hubble Frontier Fields clusters based on ICL measurements and SL-only mass models from the literature, using the same ($r=140$ kpc) threshold. Their MHD values range from $\mytilde16$ kpc to $\mytilde50$ kpc, with a mean of $\mytilde25$ kpc. In comparison, the MHD of the Bullet Cluster is relatively small.

We find that a similarly small MHD ($19.80\pm12.46\ \mathrm{kpc}$) is obtained even when increasing the threshold to $\mytilde400~ \mathrm{kpc}$, significantly expanding the area over which the distributions are compared. This supports the conclusion that the ICL is a reliable tracer of the DM distribution, even in a highly dynamic merging environment.

We note that parametric mass models adopt bright cluster galaxies as priors for the locations of dark matter halos. As a result, MHD measurements based on such models can be inherently biased toward greater similarity with the light distribution. In this regard, we highlight that our MHD analysis based on the free-form mass map is not an artifact of modeling assumptions, but a genuine morphological correspondence.

\section{Discussion} \label{sec:discussion}

\begin{figure}
    \centering
    \includegraphics[width=\linewidth]{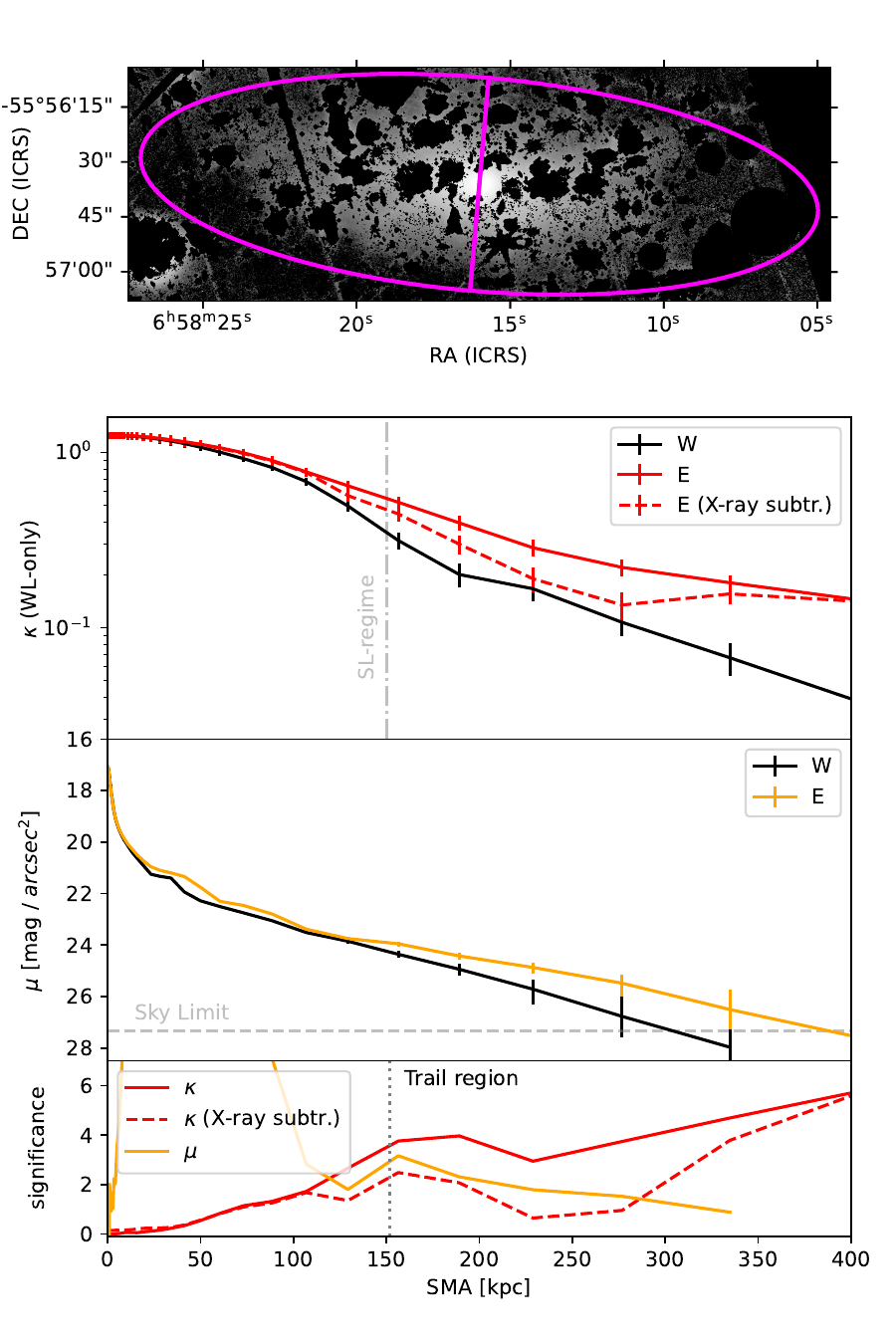}
    \caption{Radial profiles across the trail structure of mass and ICL.
    The top panel shows the masked F277W image overlaid with elliptical regions used to extract the radial profile along the extended direction.
    The bottom three panels present the radial profiles of the convergence $\kappa$, the surface brightness of BCG + ICL $\mu$, and the significance of the asymmetry between the western (black) and eastern (red/orange) sides, from top to bottom. The red dashed lines indicate the results with the gas component subtracted.
    The semi-major axis is plotted in $\mathrm{kpc}$, and error bars represent $1\sigma$ uncertainties.
    The significance shows the normalized differences between the two sides.
    A significant excess is detected on the eastern side, indicating the presence of an extended structure.}
    \label{fig:bullet_profile}
\end{figure}

\begin{figure*}
\centering
\includegraphics[width=\textwidth]{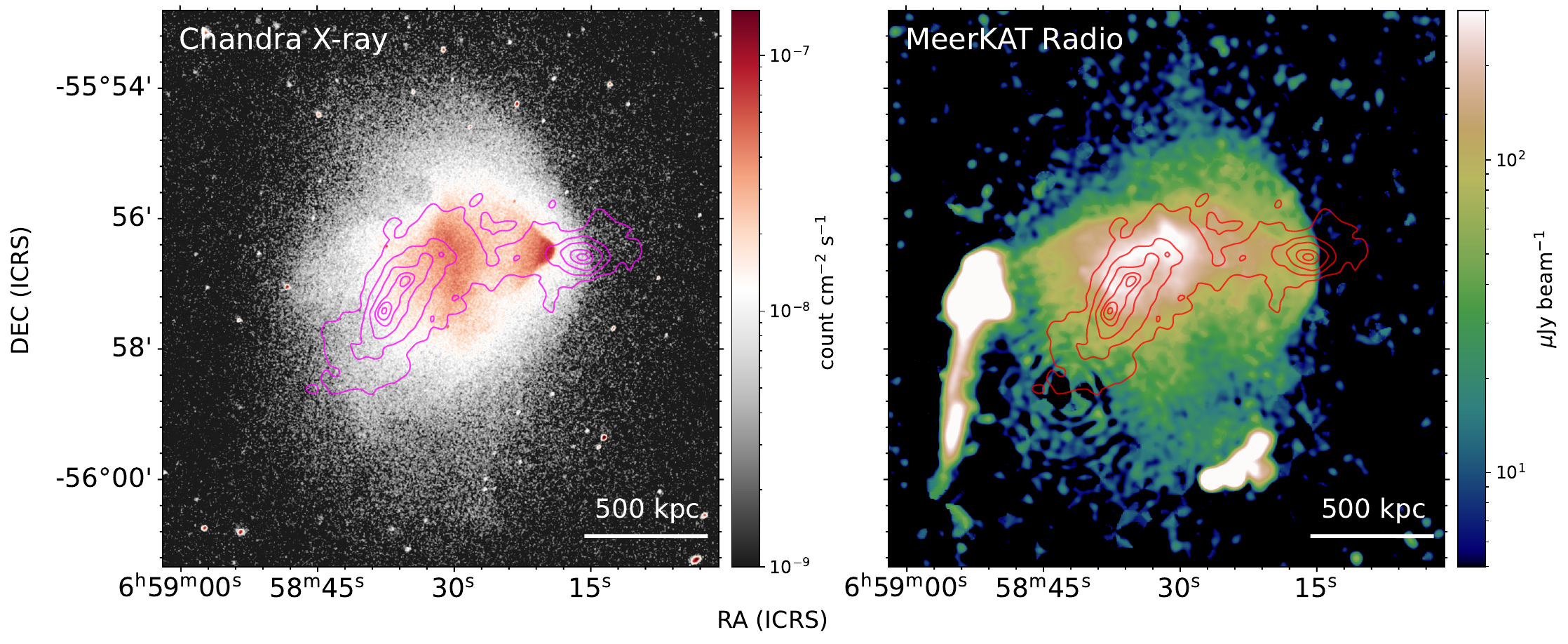} 
\caption{ICM distributions of the Bullet cluster. The left panel shows an exposure corrected 0.5-7.0 keV \textit{Chandra} X-ray map, and the right panel presents the MeerKAT 1.28 GHz radio continuum image without discrete sources. The red solid lines indicate the mass contours from our combined mass map. The contour levels correspond to $\kappa=[0.15, 0.45, 0.75, 1.05, 1.35]$.}
\label{ICM_mass}
\end{figure*}

\subsection{Mass-ICL Trail in the Subcluster}\label{subcluster_tail}
An intriguing substructure observed in both the mass and ICL maps is a trailing feature that extends from the eastern side of the subcluster toward the northern tip of the main cluster.
To assess the statistical significance of this feature, we analyze the radial profiles of the mass and ICL using the elliptical binning scheme shown in the top panel of Figure~\ref{fig:bullet_profile}, centered on the subcluster BCG. The middle two panels present the resulting profiles in mass and ICL on the eastern and western sides.

The bottom panel of Figure~\ref{fig:bullet_profile} shows the statistical significance of the difference between the two sides, computed by subtracting the western profile from the eastern one. Since the feature in the mass map is primarily constrained by the high-density JWST WL data, we estimate the mass profile uncertainties from the WL-only mass map using bootstrap resampling of the source galaxies. The uncertainties in the ICL profile account for both photon noise and background estimation errors. The integrated significance of the eastern excess is $\mytilde5\sigma$ in the mass map and $\mytilde3\sigma$ in the ICL.

We find that the mass trail feature remains statistically significant even after accounting for the contribution from the dense ICM concentration (i.e., the ``Bullet") located  $\mytilde150$~kpc east of the subcluster BCG. We reassessed the significance of the feature using two methods. In the first, we subtracted the X-ray-inferred mass of the Bullet component \citep{2006ApJ...648L.109C} from our mass profile (see red dashed lines in Figure~\ref{ICM_mass}). In the second, we measured the significance using only the profile beyond $r \gtrsim 250$~kpc, conservatively excluding the Bullet region. In both cases, the significance remains high at the $\gtrsim 3.5\sigma$ level.

This elongation may suggest the presence of a possible mass and ICL ``bridge'' between the main cluster and subcluster. Numerical simulations suggest that low-contrast mass bridges can form after pericenter passage in cluster mergers. Observational support for such features has been reported in previous studies \citep[e.g.,][]{2024ApJ...961..186C,2024arXiv240500115H}. However, caution is warranted in interpreting the trail feature as evidence of a mass bridge, despite its statistical significance, until the potential influence of intervening structures (in lensing) and contamination from PSF wings and stray light (in ICL) can be ruled out.

\subsection{Multi-Wavelength Comparison: Mass, X-ray, and Radio Emission}\label{ICM}

The Bullet Cluster is well known for its clear dissociation between the ICM and mass distributions. Figure~\ref{ICM_mass} compares our JWST-based mass map with \textit{Chandra} X-ray \citep{2002ApJ...567L..27M} and MeerKAT radio observations \citep{2022A&A...657A..56K}, which trace thermal and non-thermal plasma components, respectively. In X-rays, the subcluster’s mass peak is offset by $\mytilde150$ kpc from the ICM core (i.e., Bullet) along the collision axis. A larger offset ($\mytilde200$~kpc and $\mytilde400$~kpc from the northern and southern BCGs of the main cluster, respectively) is observed in the main cluster. 
These offsets are consistent with previous studies \citep[e.g.,][]{2004ApJ...604..596C}. However, our free-form mass reconstruction, enabled by enhanced SL and WL constraints from JWST and performed without using the cluster galaxies as informed priors,  substantially increases the statistical significance of the observed dissociations between the mass and ICM distributions.

While the offset in the subcluster is relatively easy to reproduce, simulations of the Bullet Cluster as a binary merger have difficulty accounting for the larger mass-ICM offset in the main cluster.
However, we note that the degree of dissociation in the main cluster is energy-dependent; the high-energy \textit{NuSTAR} X-ray map \citep{2014ApJ...792...48W} and Sunyaev-Zeldovich decrement maps \citep{2009ApJ...701...42H} show more extended diffuse emission toward the main cluster mass peaks than is seen in the Chandra X-ray map.
The diffuse radio halo, which is believed to be formed by turbulence-driven re-acceleration in the ICM, traces the X-ray emission in its overall extent, with its western edge coinciding with the location of the bow shock detected in X-rays \citep[see the right panel of Figure~\ref{ICM_mass},][]{2023MNRAS.518.4595S,2023A&A...674A..53B}.
However, the brightest region of the radio halo aligns more closely with the mass distribution of the main cluster than with the X-ray emission. Furthermore, there is excess radio emission between the main cluster and subcluster, coinciding with the location of the mass-ICM trail (\textsection\ref{subcluster_tail}), further supporting the presence of diffuse structure connecting the two clusters.

The multi-wavelength analysis reveals the complex merger history of the Bullet Cluster. 
While the subcluster serves as a textbook example of a dissociative merger, the main cluster shows multiple signatures of a more intricate dynamical history, including multiple mass peaks, a mass-ICL overdensity between the two BCGs, the absence of X-ray emission, and a bright radio halo that broadly follows the mass distribution. 
As mentioned above, previous numerical studies have successfully reproduced the spatial offset of the subcluster using binary cluster mergers, but have failed to account for the large dissociation observed in the main cluster \citep{2007MNRAS.380..911S, 2008MNRAS.389..967M, 2014ApJ...787..144L}. These discrepancies suggest that the Bullet Cluster is unlikely to be explained by a simple binary merger and that a revised merger scenario is required to account for its observed multi-wavelength features.

\subsection{SIDM Cross-section}\label{sidm_cross_section}

The Bullet Cluster has long been regarded as a valuable laboratory for DM studies, serving both as evidence for the existence of dark matter and as a probe for measuring its self-interaction cross-section (SICS). One approach to constraining the SICS involves utilizing the offset between the BCG and the DM centroid \citep[e.g.,][]{2008ApJ...679.1173R}, under the assumption that the DM self-interaction acts as a drag force, producing a scenario wherein the DM halo lags behind the galaxy distribution, with an offset being proportional to the SICS. 
Although we are aware that this hypothesis has been challenged by several numerical studies \citep[e.g.,][]{Kahlhoefer2013, Kim2017}, here we present the SICS upper limit using the classical mass-galaxy offset approach with our current JWST result, in order to demonstrate the constraining power enabled by JWST. We follow the formalism of \cite{2008ApJ...679.1173R}, who established a relation between the observed mass centroid and BCG offset and the SICS for the Bullet Cluster.

We compute the centroid of the subcluster by using the first moment of the convergence values within the subcluster region. The SL-only mass map provides an offset between the mass centroid and the BCG of $4.09\pm0.63$ kpc, with the uncertainty derived from 100 different realizations. In contrast, the combined SL and WL mass map yields $17.78\pm0.66$ kpc. The 68\% upper limits on the SICS are $\sigma/m \lesssim 0.2$ and $0.5{\rm ~ cm^2/g}$ for the offsets of $4.09\pm0.63$ kpc and $17.78\pm0.66$ kpc, respectively.
At face value, these results represent one of the tightest constraints on the SICS to date \citep[e.g.,][]{2015Sci...347.1462H, 2019MNRAS.488.1572H}. However, as noted earlier, caution is warranted due to several caveats, including the theoretical detectability of the mass-BCG offset \citep[e.g.,][]{Kahlhoefer2013, Kim2017}, significant uncertainties in the merger scenario \citep[e.g.,][]{Wittman2018}, unknown BCG–DM misalignments prior to the merger, and the lack of consensus on the total mass of the system \citep[e.g.,][]{2016A&A...594A.121P}. Therefore, although our offset measurements are significant, the derived SICS here is presented as an illustrative upper limit, intended to demonstrate the centroid constraining power enabled by JWST.

\section{Conclusion} \label{sec:conclusion}

We have presented the most detailed mass reconstruction of the Bullet Cluster to date using deep JWST imaging, combining 146 strong-lensing constraints with high-density (398 sources arcmin$^{-2}$) weak-lensing measurements. Employing the free-form {\tt MARS} algorithm, we constructed a mass map without relying on the light-traces-mass assumption, enabling the identification of detailed complex substructures in this dynamically active system.

Our mass model reveals a highly elongated main cluster composed of multiple mass clumps aligned with BCGs, and a more compact subcluster with a single dominant mass peak. We detect an extension on the eastern side of the subcluster in both the mass and ICL distributions, potentially tracing a low-contrast bridge between the main and subclusters. The strong spatial correlation between mass and ICL, quantified via a modified Hausdorff distance of $\sim$20 kpc, supports the utility of ICL as a luminous tracer of dark matter even in extreme merging environments.

A multi-wavelength comparison points to a more complex merger history than the classic binary scenario suggests. Using the high-resolution mass map, we also place an updated upper limit on the self-interaction cross-section of dark matter, though interpretation is limited by uncertainties in merger geometry and pre-merger conditions.

This work demonstrates the power of JWST in precisely constraining dark matter distributions and uncovering low-contrast structures previously inaccessible. Future studies incorporating wide-field imaging and numerical modeling will be essential for measuring the total mass and reconstructing the full merger history of the Bullet Cluster, advancing its role as a key laboratory for testing dark matter physics.

\begin{acknowledgments}
This work is based on observations created with NASA/ ESA Hubble Space Telescope and NASA/ESA/CSA JWST, downloaded from the Mikulski Archive for Space Telescope (MAST) at the Space Telescope Science Institute (STScI). The specific observations analyzed can be accessed via \dataset[doi: 10.17909/8zea-jv19]{http://dx.doi.org/10.17909/8zea-jv19}.
This paper employs a list of Chandra datasets, obtained by the Chandra X-ray Observatory, contained in the Chandra Data Collection ~\dataset[doi: 10.25574/cdc.373]{https://doi.org/10.25574/cdc.373}.
This research has made use of software provided by the Chandra X-ray Center (CXC) in the application packages CIAO and Sherpa.
MGCLS data products were provided by the South African Radio Astronomy Observatory and the MGCLS team and were derived from observations with the MeerKAT radio telescope. The MeerKAT telescope is operated by the South African Radio Astronomy Observatory, which is a facility of the National Research Foundation, an agency of the Department of Science and Innovation. We express our gratitude to Andrea Botteon for providing the MeerKAT radio continuum image. 
We acknowledge support for the current research from the National Research Foundation (NRF) of Korea under the programs 2022R1A2C1003130 and RS-2023-00219959 (MJJ), RS-2024-00413036 (SC), and RS-2024-00340949 (WL).
\end{acknowledgments}

\bibliographystyle{aasjournal}
\bibliography{main}

\end{document}